\documentclass[aps,prd,groupedaddress,]{revtex4}
\usepackage{latexsym}
\usepackage[dvips]{graphicx}

\newcommand{\eq}{\begin{equation}}
\newcommand{\feq}{\end{equation}}
\newcommand{\eqn}{\begin{eqnarray}}
\newcommand{\feqn}{\end{eqnarray}}
\newcommand{\arr}{\begin{eqnarray*}}
\newcommand{\farr}{\end{eqnarray*}}
\newcommand{\beq}{\begin{equation}}
\newcommand{\eeq}{\end{equation}}
\newcommand{\bea}{\begin{eqnarray}}
\newcommand{\eea}{\end{eqnarray}}

\def\beq{\begin{equation}}
\def\eeq{\end{equation}}
\def\feq{\end{equation}}
\def\bea{\begin{eqnarray}}
\def\eea{\end{eqnarray}}
\def\bc{\begin{displaymath}}
\def\ec{\end{displaymath}}
\def\lb{\label}

\def\la{\lambda}

\def\Om{\Omega}

\def\lb{\label}

\begin{document}


\title{Microscopic Entropy of Non-dilatonic Branes: a 2D Approach}

\author{Mariano Cadoni}
\email{mariano.cadoni@ca.infn.it}
\author{Nicola Serra}
\email{nicola.serra@ca.infn.it}
\affiliation{Dipartimento di Fisica,
Universit\`a di Cagliari, and INFN sezione di Cagliari, Cittadella
Universitaria 09042 Monserrato, ITALY}


\begin{abstract}
We investigate   non-dilatonic  $p$-branes in the 
near-extremal, near-horizon regime. A two-dimensional gravity model, 
obtained from 
dimensional reduction, gives an effective description of 
the brane. We show that the AdS$_{p+2}$/CFT$_{p+1}$ correspondence at 
finite temperature admits an effective description in terms of a
AdS$_{2}$/CFT$_{1}$ duality
endowed with a scalar field, 
which breaks the conformal symmetry and generates a non-vanishing 
central charge. The entropy of the CFT$_{1}$ is computed using Cardy 
formula. Fixing in a natural way  a free, dimensionless, parameter 
introduced in the 
model by a renormalization procedure,  we find exact agreement between 
the CFT$_{1}$ entropy and the Bekenstein-Hawking entropy of the brane.

\end{abstract}


\maketitle

\section{Introduction}
The brane solutions  of string and M-theory are important for several 
reasons. The non-dilatonic $p$-brane solutions play a crucial role in 
the formulation of the anti-de Sitter/Conformal Field theory 
(AdS/CFT) correspondence \cite{Maldacena:1997re,Witten:1998qj,Gubser:1998bc}. 
For  instance, in the
 3-brane case, the low energy limit of string theory is found to have 
two  different descriptions,
 each of them splitting  into two decoupled pieces. The first is free bulk 
supergravity and the near-horizon
 geometry of the   extremal 3-brane ($AdS_5 \times S^5$) and the second is free bulk 
supergravity and ${\cal N}=4$, $U(N)$ super Yang-Mills 
theory (SYM) in four dimensions. This  led Maldacena to identify  
string theory 
on  $AdS_5 \times S^5$ and 
 the SYM theory as duals. Similar arguments led Maldacena to 
propose a duality between string theory on the near-horizon
 geometry of extremal non-dilatonic $p$-branes in D-dimensions ($AdS_{p+2}\times 
S^{D-p-2}$) and a conformal field theory in $p+1$ dimensions.  

Brane solutions are also interesting from a slightly different, albeit 
related, point of view.  $p$-branes are classical solutions of 
supergravity (SUGRA) theories in $D$ dimensions. Being gravitational 
configurations, they may be endowed with an event horizon and become 
black $p$-branes. From this point of view they can be considered as 
a generalization of charged black hole solutions of general relativity.
In particular, one can associate to them a thermodynamical entropy 
using  Bekenstein-Hawking area law.
Similarly to the black hole case, one has to face the  problem of 
giving a microscopical interpretation of the Bekenstein-Hawking entropy 
of the brane.

In view of the AdS/CFT correspondence, one is tempted to use the 
CFT$_{p+1}$ dual theory to compute the microscopical entropy of the 
near-horizon,
near-extremal $p$-brane.  However, this is not so easy.
The AdS/CFT duality is assumed to hold at zero
temperature,  corresponding to the extremal brane, which has also zero 
entropy. Near-extremal branes  have non-vanishing temperature and entropy, 
but finite temperature effects break conformal invariance. If the 
AdS/CFT duality survives finite temperature effects, the near-extremal brane 
should be described by a CFT at finite temperature. 
Indications that this could be the case come 
from calculations for the 3-brane. 

Klebanov et al. compared the entropy of the 3-branes 
with  that of   finite temperature,  weak coupled 
$U(N)$ gauge theory.
 They
found an  agreement of the two results  
 up to a numerical  factor \cite{Gubser:1996de,Klebanov:1996un}.
The origin of the discrepancy factor is qualitatively well understood.
The gauge theory computation is performed  at weak 't Hofft 
coupling, whereas the gravity description is 
 assumed to be valid  at strong 't Hofft 
coupling. Also the result for the 1-brane in $D=6$ indicates that 
near-extremal branes can be described  by a finite temperature CFT.
In this latter case, the near-extremal brane can be identified  as a
Ba\~nados-Teitelboim-Zanelli (BTZ) black hole, whose entropy can be 
exactly reproduced using a two-dimensional (2D) CFT at finite temperature 
\cite{Strominger:1997eq}. 

For M-branes (the 2,5-brane) the situation looks rather different.
Here the AdS/CFT duality is of little help. The Klebanov et al. 
calculation, which uses a dual, weak-coupled, field theory,  
reproduces correctly  the scaling behavior of the brane entropy with the 
temperature but not that with the number of branes $N$. 
Thus, for the 2,5-brane we have rather weak indications that the near-extremal 
brane can admit a description in terms of a finite temperature CFT$p+1$.

Other attempts  to explain the entropy of non-extremal $p$-branes 
use a generalization of the approach proposed by Strominger and Vafa 
to compute the entropy of extremal BPS black holes \cite{Strominger:1996sh}. 
One tries to explain the entropy of the brane is terms  of 
states of the string living on the brane
\cite{Callan:1996dv,Horowitz:1996fn,Breckenridge:1996sn,Horowitz:1996ac,Maldacena:1996ya}.

Considering this situation it is worth to explore other  possibilities 
to describe   near-extremal non-dilatonic 
$p$-branes, which can be   used to give a microscopic 
interpretation of the brane entropy.
In Ref. \cite{Cadoni:2003vi} has been proposed an effective description  of 
the  near-extremal 3-brane in 
terms of a AdS$_{2}$/CFT$_{1}$ duality endowed with a scalar field 
which breaks the conformal symmetry and generates a non-vanishing 
central charge. The Bekenstein-Hawking entropy of the 3-brane could 
be matched by CFT$_{1}$ calculations up to a numerical factor.

In this paper we improve  the method used in Ref. \cite{Cadoni:2003vi} and 
generalize it to all the relevant non-dilatonic branes. 
In particular, we will be able to reproduce exactly the thermodynamical entropy 
of  the near-extremal non-dilatonic $p$-branes fixing in a natural way a dimensionless, 
 renormalization parameter, which appears as free parameter in our 
 calculations. 

In section II we briefly review some basic facts about black $p$-branes. Later on 
(section III) we perform
 a dimensional reduction of the brane  to obtain a 2D gravity model,
 which gives an effective description of the near-horizon, near-extremal brane. 
 In section IV we  study
 the group of  asymptotical symmetries (ASG) of the 2D solutions.
 A one-dimensional CFT  will emerge as dual description  of the 
 2D bulk gravity theory.
The central charge of the associated Virasoro algebra is computed in 
section V. 
Making use of a renormalization procedure, we will be able to 
find  finite charges associated with the  ASG, at the price of
introducing a dimensionless free parameter $\beta$. 
The entropy of the $p$-brane is then calculated via the Cardy formula. Fixing the 
  parameter $\beta$  
we  reproduce exactly the 
thermodynamical entropy of the brane.

\section{Non-dilatonic black $p$-Branes}

In this section we briefly review some well-known facts about 
$p$-branes.
Black $p$-branes are classical Ramond-Ramond (RR) charged solutions 
of SUGRA theory  in $D$ dimensions 
\cite{Horowitz:cd,Duff:1991pe,Duff:1994an,Gregory:1995qh,Stelle:1996tz,Peet:1997es}.
They can be also considered as the low energy limit of string and 
M-theory.

In  the Einstein frame the bosonic part of the action reads:

\beq \label{azione}
A= \frac{1}{2k_D^2}\int d^Dx\sqrt{-g}\left( 
R-\frac{1}{2}(\nabla\phi)^2-\frac{1}{2n!}F_n^2e^{a \phi}\right),
\feq
where $\phi$ is the 
dilaton, 
$F_n$ is the field strength of an $(n-1)$-form potential 
$F_n=dA_{n-1}$, and $a$ is a constant depending on the dimensional 
reduction that produces the action in $D$ dimensions.  
The metric part of the electric  solution  of the action 
(\ref{azione})  
is   
\cite{Gibbons:1994vm,Petersen:1999zh,Aharony:1999ti}
(the magnetic solution is obtained 
using Hodge duality) 

\eqn\lb{brane}
ds^2&=&[H(r)]^{-\frac{2(d-2)}{\delta}}\left(-f(r)dt^2+
\sum_{i=1}^{p}dx_idx^i\right)+
[H(r)]^{\frac{2(p+1)}{\delta}}\left(f^{-1}(r)dr^2+r^2d\Om_{d-1}^2\right)
\nonumber\\
H(r)&=&1+\left(\frac{h_p}{r}\right)^{d-2}, \quad 
f(r)=1-\left(\frac{r_0}{r}\right)^{d-2} \quad 
D=d+p+1,\nonumber\\
\delta&=&(p+1)(d-2)+\frac{a^2(D-2)}{2},\quad
h_p^{2(d-2)}+r_0^{d-2}h_p^{d-2}=\frac{\delta Q^2}{2(d-2)^2(D-2)},
\feqn
where $Q$ is the RR charge, $r_0$ and $h_p$ are integration constants
related to  the mass and charge 
of the brane. 
The solution (\ref {brane}) can be regarded as a generalization of the 
Reissner-Nordstrom charged black hole solution,
for this reason  it  is called black $p$-brane.

We are interested in non-dilatonic branes: M-branes 
(the 2-brane and 5-brane in 
eleven dimensions) and dyonic  branes  with equal magnetic ed 
electric charges in   $D=2p+4$, with $p$ odd
(1-brane in $D=6$ and  3-brane in $D=10$). 
M-branes are non-dilatonic simply because there 
is no dilaton in eleven 
dimensions. Dyonic branes with equal charges in   $D=2p+4$ and $p$ odd
are  self-dual 
solutions of theories 
with  self dual $(p+2)$-field strengths. They are characterized by
a constant dilaton and can be thought of  as intrinsically non-dilatonic.
In what follows we will consider only non-dilatonic branes.

In the extremal limit 
($r_0=0$) the
 metric (\ref{brane}) becomes:
\eqn\lb{ebrane}
ds^2&=&[H(r)]^{-\frac{2}{p+1}}\left(-dt^2+\sum_{i=1}^p dx_idx^i\right)+
 [H(r)]^{\frac{2}{d-2}}(dr^2+r^2d\Om_{d-1}^2)\nonumber\\
h_p^{d-2}&=&\frac{Q}{\sqrt{\alpha}(d-2)}, \quad 
\alpha=\frac{2(D-2)}{(d-2)(p+1)}.
\feqn
The brane tension $T_p$ and the $D$-dimensional Newton constant 
$G_{D}$,  can be 
written in terms of  
the $D$-dimensional Planck length $l_{D}$¥ and the string coupling 
constant $g_s$,

\beq\lb{e1}
T_p=\frac{2\pi}{(2\pi l_D)^{p+1}g_s} , \quad 2k_D^2=16\pi 
G_D=\frac{(2\pi l_D)^{D-2}g_s^2}{2\pi}.
\feq

We can also define,

\beq\lb{e2}
e_p=\frac{1}{\sqrt{2}k_D}\int_{S^{d-1}} 
\phantom{}^{*}F_{d-1}=\frac{Q\Om_{d-1}}{\sqrt{2}k_D}.
\feq

For a single $p$-brane the flux of the RR field is $\sqrt{2}k_DT_p$. 
If we consider
 $N$ coincident $p$-branes we have
\beq\lb{e10}
N=\frac{e_p}{\sqrt{2}k_DT_p}=\frac{\Om_{d-1}Q}{2k_D^2 T_p}, \quad 
h_p^{d-2}=\frac{(2\pi l_D)^{d-2}Ng_s}{\sqrt{\alpha}(d-2)\Om_{d-1}}.
\feq

The AdS/CFT duality arises considering  the near-horizon limit,
\beq\lb{nhl}
(r,l_D)\rightarrow 0\quad \mbox{keeping fixed}\quad u^{(p+1)/(d-2)}=
r l_{D}^{-\alpha(p+1)/2}
\feq
of the 
brane solution. In this limit the extremal brane (\ref{ebrane})
becomes 
\beq\label{Nino}
ds^2=\la_p^2u^2\left( -dt^2+\sum_{i=1}^p 
dx_idx^i\right)+\frac{du^2}{\la_p^2u^2}+R_{0p}^2d\Om_{d-1}^2,\quad
\feq
where $R_{0p}=h_p$ and $\lambda_p=(d-2)/[(p+1)R_{0p}]$.
The near-horizon region has the $AdS_{p+2}\times S^{d-1}$ 
geometry.
 This was the starting point of the  Maldacena conjecture 
 about a  duality between 
 string  theory 
in this background 
and  conformal field theory in $(p+1)$ dimensions.  The 
isometry group of 
 $AdS_{p+2}$ is identified with the conformal symmetry of 
$(p+1)$-dimensional Minkowski space.

Excitations above extremality  break the conformal symmetry and the 
brane acquires finite temperature and entropy. 
The near-horizon, near-extremal form of the solution can be easily 
obtained by taking in Eq. 
(\ref{brane}) the near-horizon limit (\ref{nhl})  
and the energy above extremality finite.
One finds
\beq\label{EnricoFois}
ds^2=\la_p^2u^2\left[ -\left(1-(\frac{u_0}{u})^{p+1}\right)dt^2+\sum_{i=1}^p 
dx_idx^i\right]+\frac{du^2}{\la_p^2u^2\left[1-{(\frac{u_0}{u})}^{p+1}\right]}+
R_{0p}^2d\Om_{d-1}^2,
\feq
where $u$ and  $u_{0}$ are  defined as in Eq. (\ref{nhl}). 

Using the Bekenstein-Hawking formula, we can easily calculate the 
entropy of the  brane (\ref{EnricoFois}).  Working in  
the canonical ensemble, we can express the entropy $S_{p}$ 
as a function of the temperature $T$ and volume $V$ of the brane,
\beq\lb{entropy}
S_{p}¥=a_{p}V T^{p},
\feq
where $a_{p}$ depends on the number $N$ of coincident branes and is 
given for the 1,2,3,5-brane under consideration as follows,
\beq\lb{a}
a_{1}=\pi N^2,\quad 
a_{2}=\frac{2^{\frac{7}{2}}\pi^2}{27}N^{\frac{3}{2}},
\quad a_{3}=\frac{\pi^2}{2}N^2, \quad a_{5}=\frac{2^7}{3^6}\pi^3 N^3.
\feq
Klebanov et al  tried to give a 
microscopic interpretation  of the thermodynamical entropy of the brane 
using a system of weak interacting   brane 
excitations \cite{Gubser:1996de,Klebanov:1996un}. They could 
reproduce the scaling behavior (\ref{entropy}), but  they found 
an expression   for the coefficients $a_{p}$ depending on a 
parameter $\tilde{n}$, which characterizes the field content of the 
model:
\beq\lb{a1}
a_{1}=\frac{\pi}{2}\tilde{n},
\quad a_{2}=\frac{7}{8\pi}\zeta(3)\tilde{n},
\quad a_{3}=\frac{\pi^2}{12}\tilde{n}, \quad 
a_{5}=\frac{\pi^3}{40}\tilde{n}.
\feq
    
For the 3- and 1-brane the AdS/CFT correspondence allows an easy 
identification of  parameter $\tilde n$. 
The CFT$_4$ dual to AdS$_{5}$ is well known, it is   
${\cal{N}} = 4,$ $U(N)$ SYM. 
This fact enables us  to identify $\tilde{n}=8N^2$ 
\cite{Gubser:1996de}.
In this way the statistical entropy is in 
 agreement with the thermodynamical one up to a $3/4$ 
factor. The origin of the discrepancy factor is well understood.
The gauge theory computation is performed at zero 't Hofft coupling, 
whereas the gravity description is assumed to be valid  in the strong 
coupling regime \cite{Gubser:1998nz, Klebanov:2000me}.
For the 1-brane the AdS/CFT correspondence allows us to reproduce  
exactly the thermodynamical result (\ref{a}).
In this case, neglecting the 3-sphere of constant radius,  
the brane solution (\ref{EnricoFois})  is nothing but
the Ba\~nados-Teitelboim-Zanelli (BTZ) black hole, whose 
microscopic entropy has been calculated by  Strominger \cite{Strominger:1997eq}, 
leading to  
the identification 
$\tilde{n}=2N^{2}$ 

For the two M-branes the situation is more involved.
The AdS/CFT correspondence is here of little help, because the 
dualities 
AdS$_{4}$/CFT$_{3}$ and AdS$_{7}$/CFT$_{6}$ are poorly understood.
Moreover, in order to explain  the dependence on $N$ in Eq. (\ref{a})
we need a behavior $\tilde{n}\sim N^{3/2}$ and $\tilde{n}\sim 
N^{3}$, respectively for the 2- and 5 brane,  which is very hard to 
achieve using a  field theory. In spite of some progress, 
achieved considering  a D-brane-D-antibrane system 
\cite{Danielsson:2001xe,Danielsson:2002qg,KalyanaRama:2004fk,Saremi:2004pi,Halyo:2004bx},
 this still remains a puzzling point,
which is related with our lack of knowledge about M-theory.

In this paper we will use a 2D approach to the problem  of giving a 
microscopical interpretation  for the entropy of non-dilatonic branes.
The first step in this direction is to perform a dimensional reduction
in order to obtain a 2D effective description of the brane. This will 
be the subject of the  next section.

\section{Dimensional reduction}

The near-horizon,   near-extremal non-dilatonic brane solution
(\ref{EnricoFois}) factorizes as 
direct product of a $(p+2)$-dimensional spacetime, which is
asymptotically AdS$_{p+2}$, and   a $(d-1)$-sphere $S^{d-1}$
of constant radius.
This fact allow us to derive, by dimensional reduction,  
a 2D effective  gravity model, which describes 
the spherically symmetric excitations of the brane 
above extremality. 
We can perform the dimensional reduction from $D$ to two dimensions 
 using  
the ansatz:

\beq\label{Carla}
ds^2_{D}¥=ds^2_2+\phi^{\frac{2}{p}}\sum_{i=1}^p 
dx_idx^i+R_{0p}^2d\Omega_{d-1}^2,
\feq
where $\phi$ is  a scalar field , which parametrizes the volume 
$ \cal{V}$ of the brane embedded in the $(p+2)$-dimensional spacetime,
\beq\label{Giovanni}
{\cal{V}}=\phi V.
\feq
For the RR field strength we have $F_n^2/n!=Q^2/h_p^{2(d-1)}$.
Performing the dimensional reduction in the $D$-dimensional action
(\ref{azione})
we get the 2D effective model,
\beq\label{Rossella}
A_{2D}=k\int d^2x \sqrt{-g}\phi \left\{ R+\left(\frac{p-1}{p}\right) 
\frac{(\nabla \phi)^2}{\phi^2}+\Lambda\right\},
\feq
where the cosmological constant $\Lambda=R_{(d-1)}-(F_n^2/2n!)$,
$R_{(d-1)}$ being 
the  scalar curvature of $S^{d-1}$ and the constant $k$ is
\beq\label{Martedi4Maggio2004}
k=\frac{\Omega_{d-1}R_{0p}^{d-1}V}{2k_D^2}=\frac{2\pi N 
VR_{0p}}{\sqrt{\alpha}(d-2)(2\pi l_D)^{p+1}g_s}.
\feq
The class of 2D gravity models described by the action 
(\ref{Rossella}) has been already investigated  in the literature 
\cite{Cadoni:2001ew,Youm:1999xn,Medved:2003ar}. In particular,  they admit 
the, asymptotically AdS, 
2D black hole solutions,
\beq\label{MartaDeidda}
ds^2=-(b^2r^2-A^2(br)^{1-p})+\frac{dr^2}{b^2r^2-A^2(br)^{1-p}},\quad
\phi=\phi_0(br)^{p},
\feq
where $b^2=\Lambda/[p(1+p)]$ and  $\phi_{0},A$ 
are integration constants.
The thermodynamical behavior of the 2D black hole  
is characterized by  mass $m_{bh}$, temperature $T_{bh}¥$ and entropy
$S_{bh}$,

\beq\label{MTS}
m_{bh}=\frac{p}{2}\phi_0 A^2 b, \quad 
T_{bh}¥=\frac{b(p+1)}{4\pi}A^{\frac{2}{p+1}}, 
\quad S_{bh}=2\pi \Phi_{0}¥A^{\frac{2p}{p+1}}.
\feq

The 2D black hole solution (\ref{MartaDeidda}) gives an 
effective description of the $D$-dimensional 
brane solution (\ref{EnricoFois}).  
The 2D integrations constant $A$ and $\phi_0$ can be identified  in 
terms of the physical 
parameters of the brane. Comparing Eq.(\ref{EnricoFois}) with 
Eq. (\ref{MartaDeidda}), we can easily see  
that  $A$ is related to energy of  the brane  excitations 
above extremality:
\beq
A^2=\lambda_p^{p+1}u_0^{p+1}.
\feq
We can fix the value of  $\phi_0$ using a (classical) scale 
symmetry of the 2D action (\ref{Rossella}).
Rescaling the 
 scalar field $\phi \rightarrow \mu \phi$ the action changes as 
$A_{2D}\rightarrow \mu A_{2D}$. 
In this way we can change the normalization factor in front of the 
action. Choosing the normalization of Ref. \cite{Cadoni:2001ew} 
($k=1/2$) we have  $\phi_0=2k$. 
One can easily check that the thermodynamical behavior of the 2D black 
hole 
solution reproduces 
exactly the  thermodynamics of the near-extremal brane 
(\ref{EnricoFois}). In fact, we have  $S_{brane}=S_{bh}$,
 $E_{brane}=m_{bh}$ and $T_{brane}=T_{bh}$, where $E_{brane}$ is
the  energy of a brane excitation above extremality. This fact has a   
natural interpretation.  The thermodynamics of the brane 
is determined by the  behavior on the horizon  of the 2D $(r,t)$ section 
of  metric (\ref{EnricoFois}), which is exactly given  by the 2D black hole.

On the other hand both the non-dilatonic brane and the 2D black hole
seem to have a dual descriptions in terms of a conformal field theory.
For the brane the dual theory is a CFT$_{p+1}$, whereas for the 2D black 
hole it is  a CFT$_{1}$ \cite{Cadoni:2001ew,
Cadoni:1998sg,Cadoni:1999ja,Cadoni:2000ah,
Cadoni:2000gm, Caldarelli:2000xk, Astorino:2002bj}.
 This gives  a strong indication that the 
$CFT_1$ can be used both to give an
effective description of the $CFT_{p+1}$ and   
to obtain a microscopic derivation of the entropy of the brane,
in agreement with the philosophy of the 
holographic principle.
To achieve this goal we first need to investigate the asymptotical 
symmetries of the 2D solutions, which will be the topic discussed in 
the next section.

\section{Asymptotical symmetries}
The group of asymptotical symmetry (ASG) of the metric 
(\ref{MartaDeidda}) is the group of transformations which leaves 
the asymptotic, $r\to\infty$, behavior of the metric invariant.
The case $p=1$ is well known, it corresponds to 2D anti-de Sitter 
spacetime
(AdS$_{2}$). Its ASG  was investigated
in various papers and the problem of the microscopical explanation
of the entropy of the corresponding 2D black hole has been completely 
solved
\cite{Cadoni:1998sg,Cadoni:1999ja,Caldarelli:2000xk,Cadoni:2000ah,
Cadoni:2000gm}. 
In this paper we  focus our attention on the other three cases 
($p=2,3,5$).
The asymptotical symmetries of the metric  (\ref{Rossella}) were 
investigated in Ref.  \cite{Cadoni:2001ew}.
In that paper the ASG was identified with the group of 
reparametrizations of the 1-dimensional, $r\to\infty$, timelike 
boundary of the AdS$_{2}$ spacetime  (the $diff_1$ group). 
It was shown that
the generators of the group
satisfy a Virasoro algebra. Unfortunately, the charges  associated 
with the generators and the central extension of the Virasoro algebra
were  found to be divergent. A renormalization 
 procedure was not applicable directly because it erases identically 
the charges. The divergence 
is due to the power behavior $\phi\sim r^p$ of the scalar field for 
$r\rightarrow \infty$. 
In this paper we  propose a general method 
to renormalize the charges in a consistent way. Our method is a
generalization of the renormalization procedure proposed in 
Ref. \cite{Cadoni:2003vi}.

We can separate 
a finite from a divergent part  
in the  charges  performing the change of coordinate:
\beq\label{Martina}
(br)^{p-1}\rightarrow (br)^{p-1}+\beta A^{\frac{2(p-1)}{p+1}},
\feq
where $\beta$ is an arbitrary  dimensionless  
parameter, whose value cannot be 
fixed by the renormalization procedure.
Using scale and dimensional arguments one can easily understand that 
Eq. 
(\ref{Martina}) is 
the most general translation of the quantity $(br)^{p-1}$.
An uniform treatment of both even and odd branes is not possible.
In the following we will distinguish the two cases. 

Performing the change of coordinate
 (\ref{Martina}) and expanding the solution (\ref{MartaDeidda})
 near $r\rightarrow 
\infty$ we have for $p=2$,

\bea\label{Gesuina}
                       &g_{tt}&=-b^2r^2-2\beta 
A^{\frac{2}{3}}br-\beta^2A^{\frac{4}{3}}+\frac{A^2}{br}+O[r^{-2}], 
\nonumber\\
            &g_{rr}&=\frac{1}{b^2r^2}-\frac{2\beta 
A^{\frac{2}{3}}}{b^3r^3}+\frac{3\beta^2A^{\frac{4}{3}}}{b^4r^4}+
\frac{(1-4\beta^3)A^2}{b^5r^5}+O[r^{-6}],\nonumber \\
            &\phi&=\phi_0(b^2r^2 +2\beta 
A^{\frac{2}{3}}(br)+\beta^2A^{\frac{4}{3}}),  
\eea
whereas for $p=3,5$ we get
\bea\label{Gesuino}
             &g_{tt}&=-b^2r^2-\frac{2}{(p-1)}\beta 
A^{\frac{2(p-1)}{p+1}}(br)^{3-p}+A^2(br)^{1-p}-\frac{3-p}
{(p-1)^2}\beta^2A^{\frac{4(p-1)}{p+1}}(br)^{-(p+1)}+O[r^{2(1-p)}],
\nonumber\\
            &g_{rr}&=\frac{1}{b^2r^2}-\frac{2\beta 
A^{\frac{2(p-1)}{p+1}}}{(br)^{p+1}}+\frac{A^2}{(br)^{p+3}}+
\frac{3\beta^2A^{\frac{4(p-1)}{p+1}}}{(br)^{2p}}+O[r^{-2(p+1)}],
\nonumber \\
     &\phi&=\phi_0(b^pr^p +\frac{p}{p-1}\beta 
A^{\frac{2(p-1)}{p+1}}(br)+\frac{p}{2(p-1)^2}
\beta^2A^{\frac{4(p-1)}{p+1}}
(br)^{2-p})+O[r^{3-2p}].         
\eea
In both cases the metric is asymptotically AdS. 
In view of Eqs. (\ref{Gesuina}), (\ref {Gesuino}), we are led to impose 
the following boundary conditions,
\bea\lb{bc}
&g_{tt}&=-b^2r^2-2\beta A^{\frac{2}{3}}br+\gamma_{tt}+\frac{\Gamma_{tt}}{br}+O[r^{-2}], 
\nonumber \\
&g_{rr}&=\frac{1}{b^2r^2}-\frac{2\beta 
A^{\frac{2}{3}}}{b^3r^3}+\frac{\gamma_{rr}}{b^4r^4}+\frac
{\Gamma_{rr}}{b^5r^5}+O[r^{-6}],\nonumber \\
&g_{rt}&=\frac{\gamma_{rt}}{b^3r^3}+O[r^{-4}], \nonumber \\
&\phi&=\phi_0(\rho b^2r^2+2\rho \beta A^{\frac{2}{3}} 
br+\gamma_{\phi\phi}+\frac{\Gamma_{\phi\phi}}{br}+O[r^{-2}]), 
\eea
for $p=2$, whereas for $p=3,5$ we have

\bea\lb{bcc}
&g_{tt}&=-b^2r^2+\gamma_{tt}+\frac{\Gamma_{tt}}{b^2r^2}+
\frac{\theta_{tt}}{b^4r^4}+O[r^{-6}], 
\nonumber \\
&g_{rr}&=\frac{1}{b^2r^2}+\frac{\gamma_{rr}}{b^4r^4}+
\frac{\Gamma_{rr}}{b^6r^6}+\frac{\theta_{rr}}{b^8r^8}+O[r^{-10}],
\nonumber \\
&g_{rt}&=\frac{\gamma_{rt}}{b^3r^3}+\frac{\Gamma_{rt}}
{b^5r^5}+O[r^{-7}],\nonumber\\
&\phi&=\phi_0(\rho 
(br)^p+\gamma_{\phi\phi}(br)^{p-2}+\Gamma_{\phi\phi}(br)^{p-4}+
\theta_{\phi\phi}(br)^{p-6}+O[r^{p-8}])
\eea
In Eqs. (\ref{bc}) and (\ref{bcc}), $\rho$, $\gamma$, 
$\Gamma$, $\theta$ are boundary fields, which depend only on the 
coordinate $t$ and describe 
deformations of the 
metric and of the scalar $\phi$. 

The Killing vectors that preserve the boundary conditions (\ref{bc}) 
and define the ASG, are 
\beq\lb{kl}
\chi^r=-\stackrel{.}{\varepsilon}(t)(r+\frac{\beta 
A^{\frac{2}{3}}}{b})+O[r^{-1}], \quad 
\chi^t=\varepsilon(t)+\frac{\stackrel{..}{\varepsilon}(t)}{2b^4r^2}
(1-\frac{2\beta 
A^{\frac{2}{3}}}{br})+O[r^{-4}],
\feq
those that preserve the boundary conditions (\ref{bcc}) are instead,
\beq\label{Grazia}
\chi^r=-\stackrel{.}{\varepsilon}(t)r+O[r^{-2}], \quad 
\chi^t=\varepsilon 
(t)+\frac{\stackrel{..}{\varepsilon}(t)}{2b^4r^2}+O[r^{-4}],
\feq
 where $\varepsilon(t)$ is an arbitrary function of time and the 
dot  
denotes derivative with respect to time.
Notice that for $p=2$ both the boundary conditions (\ref{bc}) and 
the Killing vectors (\ref{kl}) depend on the parameters $A,\beta$.
This dependence is consequence of the coordinate transformation 
(\ref{Martina}).
For $p$ odd, there are contributions to the unnormalized charge 
coming from terms of order higher than $r^{-4}$ in Eq. 
(\ref{Grazia}). However,  these terms do not contribute 
to the renormalized
 charges.  We can consistently neglect them. 
 The generators $L_n$ of 
the ASG span
a Virasoro Algebra:
\beq\label{FrancescaRiccia}
[L_n, L_m]=(n-m)L_{n+m}+\frac{c}{12}(n^3-n)\delta_{n+m,0},
\feq
where we allow for a non-vanishing central charge $c$.
We can therefore  identify the ASG as the $diff_1$ group, 
the conformal group in one dimension. 
The $diff_1$ group is an asymptotical symmetry only for the metric 
part of the solution (\ref{MartaDeidda}). The ASG  of the metric is broken
by the non-constant solution for the  the scalar $\phi$.
This breaking of the conformal symmetry  is the source 
of a non-vanishing central
 charge in the Virasoro algebra (\ref{FrancescaRiccia}) \cite{Cadoni:2000ah}.
 Moreover, we will see
  in the following  that the power law  behavior $\phi\sim r^{p}$ of 
the 
scalar field is also related to
 the appearance of the divergences in the charges associated with
 the ASG. 
 
The boundary fields $\rho$, $\gamma$, 
$\Gamma$, $\theta$ 
transform under  the action of the $diff_1$ group 
as conformal fields of  definite weight. The only 
boundary field that contributes
 to the renormalized central charge is $\rho$, whose transformation 
law is:
\beq\lb{f3}
\delta\rho= 
\varepsilon\stackrel{.}{\rho}-p\stackrel{.}{\varepsilon}\rho.
\feq
The extremal brane (\ref{Nino}) 
has the  AdS$_{p+2}\times S^{¥d-1}$ geometry and is mapped by 
the dimensional reduction 
into the AdS$_2$ spacetime, which is given  by Eq. (\ref{MartaDeidda})
with $A=0$. 
Neglecting the scalar field $\phi$, we see that   the  isometry group   
of AdS$_{p+2}$, the group $SO(2,p+1)$, locally isomorphic to 
the conformal group in $(p+1)$ dimensions, is mapped by the 
dimensional reduction into  the ASG of AdS$_2$, namely the 
$diff_1$ conformal group. 
The non-constant configuration for the scalar $\phi$  breaks 
the conformal symmetry and generates a 
non-vanishing central charge in the Virasoro algebra.
The dimensional reduction allows us to find an effective description
of the $AdS_{p+2}/CFT_{p+1}$ duality at finite temperature
in terms of a $AdS_{2}/CFT_1$ duality  with the conformal  
symmetry broken by  the scalar field $\phi$.

\section{Central charge and entropy}

Because the dual theory of the effective 2D gravity theory is an 
one-dimensional
CFT,  
 knowledge of 
the central charge in the Virasoro Algebra, allows us to calculate the entropy
of the 2D black hole (hence of the near-extremal brane) via the Cardy 
formula. We can compute the 
 central charge appearing in the Virasoro algebra 
(\ref{FrancescaRiccia}) using a canonical
 realization of the ASG. 
 
 The gravitational Hamiltonian $H$ is easily 
computed using the ADM parametrization of the metric :
\beq
ds^2=-N^2dt^2+\sigma^2(dr+N^rdt)^2,
\feq
where $N$ and $N^r$ are respectively the lapse and shift functions. 
According with the 
 Regge-Teitelboim procedure \cite{Teitelboim:1972vw,Regge:1974zd,Benguria:1976in}
 we must add surface terms $J$ to $H$, 
 needed 
to obtain well-defined 
 variational derivatives.  In the case under consideration we obtain:
\beq\lb{e4}
\delta 
J=-lim_{r\to\infty}\left\{N(\sigma^{-1}\delta\phi'-\sigma^{-2}
\phi'\delta\sigma-\frac{p-1}{p}\sigma^{-1}\phi^{-1}\phi'\delta\phi)-
N'\sigma^{-1}\delta\phi+N^r(\Pi_{\phi}\delta\phi-\sigma\delta
\Pi_{\sigma})\right\},
\feq
where $\Pi_{\phi}$ and $\Pi_{\sigma}$ are respectively the momenta 
conjugate to $\phi$ and $\sigma$.
The "orthogonality problem" \cite{Carlip:1999cy, Carlip:2002be} 
typical of two dimensions, 
can be solved introducing the 
time-integrated charges \cite{Cadoni:1998sg}:
\beq\label{Alessandra}
\widehat{J}=\frac{b}{2\pi}\int_0^{\frac{2\pi}{b}} Jdt.
\feq
The central charge $c$ can be computed using the commutator
\beq \lb{e3}
\delta_{\omega}\widehat{J}(\varepsilon)=[\widehat{J}(\varepsilon),\widehat{J}(\omega)]
\feq
However, in our case the time-integrated charges (\ref{Alessandra}) are 
divergent and the final outcome of 
the calculation is a divergent central charge \cite{Cadoni:2001ew}. A renormalization procedure  
is needed in order to 
have finite charges. 

After some manipulations we  can write Eq.(\ref{e4}) in the following form,
\beq\label{ChiaraSerra}
\delta J= \delta J_I+\varepsilon \delta M,
\feq                                                             
where $M$ is  the charge associated with time translations ($\varepsilon=1$), 
while $\delta 
J_I$ is a complicate 
function of the boundary fields and of $\varepsilon$, which for shortness 
we do not quote here.
Eq. (\ref{ChiaraSerra}) presents several divergences. 
The mass term $\delta M$  is divergent 
 because arbitrary  excitations of the  boundary fields  have 
infinite energy \cite{Cadoni:2001ew}. 
We can eliminate this kind of  divergences  considering deformations of the 
boundary fields near the classical solution 
 (on-shell deformations). This can be done using appropriately the equations of 
motion induced 
on the boundary by the equation of motion for the bulk degrees of freedom.
For $p=2$ we  make use of the following boundary  equation of motion
\beq\label{Lina}
\frac{\stackrel{.}{\rho}^2}{4b^2\rho}=5\beta^2 
A^{\frac{4}{3}}\rho-\gamma_{rr}\rho-2\gamma_{\phi\phi},
\feq
while for $p$ odd we  use,
\bea\label{Camilla}
&& 
\frac{\stackrel{.}{\rho}^2}{pb^2\rho}+p\rho\gamma_{rr}+4\gamma_{\phi\phi}=0,
\nonumber\\ 
 &&p\rho b^2\gamma_{rr}^2-(p-3)\stackrel{.}{\rho}b\gamma_{rt}+\frac{p-1}{2p}
\frac{\stackrel{.}{\rho}^2}{\rho}\gamma_{tt}-
\frac{p-1}{2p}\left(\frac{\stackrel{.}{\rho}}{\rho}\right)^2
\gamma_{\phi\phi}+\nonumber\\
&&+\frac{2(p-1)}{p}b^2\frac{\gamma_{\phi\phi}^2}{\rho}+\frac{p-1}{2p}
\frac{\gamma_{rr}\stackrel{.}
{\rho}^2}{\rho}+\frac{15(p-1)^2}{4p}b^2\gamma_{rr}\gamma_{\phi\phi}+
\frac{(p-3)p}{2}b^2\rho\Gamma_{rr}+\nonumber\\
&&+(p-3)(p-1)b^2\Gamma_{\phi\phi}+\frac{\stackrel{.}
{\rho}\stackrel{.}{\gamma}_{rr}}{2}+\frac{p-1}{p}
\frac{\stackrel{.}{\rho}\stackrel{.}{\gamma}_{\phi\phi}}{\rho}=0.
\eea
Eq. (\ref{Lina}) is obtained from the   leading  and Eqs. (\ref{Camilla}) 
from the leading and
subleading, terms  in the $r=\infty$ expansion
of the bulk field equations coming from the $g_{\mu\nu}$ variation of the  
action (\ref{Rossella}).
Taking the  variation of Eqs. 
(\ref{Lina}, \ref{Camilla}) evaluated  on the classical solution 
(\ref{Gesuina}, \ref{Gesuino}), respectively, one finds that the divergent terms  
in the mass term of Eq. (\ref{ChiaraSerra}) vanish. 
Moreover, the finite part of $M$ is 
equal to the mass $m$ of the 
solution calculated using the prescription of Ref. \cite{Mann:1992yv}.

This is not the end of the 
story.  
The term $\delta J_I$ in Eq. (\ref{ChiaraSerra}) contains also divergent parts. 
The presence of these divergences can be traced back to the large $r$ 
behavior of the scalar
field, $\phi\sim r^p$. Because this behavior is shared by all the classical 
solutions of the 2D  bulk 
theory, the most natural way to remove the divergences is to 
subtract the contribution of the 
massless background solution ($A=0$  in Eqs. (\ref{Gesuina}), 
(\ref{Gesuino})), 
\beq\label{Pippo}          
ds^2=-b^2r^2 dt^2+\frac{dr^2}{b^2r^2},\quad
\phi=\phi_0 b^pr^p.
\feq
Indicating with $J_{bg}$ the charges obtained evaluating Eq. (\ref{e4})
on the massless background and defining the renormalized charges $J_R= J-J_{bg}$, 
we get respectively for  $p$ even and odd,
\bea\lb{e6}
\delta J_R&=& \frac{\phi_0\beta 
A^{\frac{2}{3}}}{b}\left(\stackrel{.}{\varepsilon}\delta\stackrel{.}{\rho}
-\stackrel{..}{\varepsilon}\delta\rho\right)+\varepsilon \delta m,\nonumber\\
\delta J_R&=& -\frac{\phi_0 \beta 
A^{\frac{2(p-1)}{p+1}}p}{b}\left( \frac{p-2}{2^{p-3}}
\stackrel{..}{\varepsilon}\delta\rho+\frac{p-1}{2^{p-1}}\stackrel{.}
{\varepsilon}\delta\stackrel{.}{\rho}\right)+\varepsilon \delta m.
\eea
Notice that we use here a  renormalization prescription that is slightly different
from that used  in Ref. \cite{Cadoni:2003vi}. In that paper  the  charges have been
renormalized subtracting only their divergent part. 
Here, we have chosen a more natural procedure, which in general  gives a different
finite result for the charges. 
We can recover the results of Ref. \cite{Cadoni:2003vi} for the entropy of the 3-brane, by fixing 
appropriately the value
of the  renormalization parameter
$\beta$.

Taking into account that the time-integrated charges are defined only 
up to a total time derivative, we can integrate the variations $\delta 
J_{R}$ in  Eqs. (\ref{e6}) to obtain,
\beq\label{Irene}
J_R[\varepsilon] =-\frac{p\phi_0 \beta A^{\frac{2(p-1)}{p+1}}}{2^n 
b}\varepsilon \stackrel{..}{\rho}.
\feq
where  $n=(0,1)$ respectively, for $p=$(even,odd). 
The term proportional to $m$ in Eq. (\ref{e6}) has been 
canceled  by choosing appropriately the integration constant.  
$J_{R}(\varepsilon)$ in Eq. (\ref{Irene}) is related to the  energy-momentum tensor 
$T_{tt}$ of the  one-dimensional CFT, $J_{R}(\varepsilon)=\varepsilon T_{tt}$. 
Using the conformal transformation of the field $\rho$ given in Eq. 
(\ref{f3}), expanding in 
Fourier modes and using Eqs. (\ref{FrancescaRiccia},\ref{e3}),
near the classical $\rho=1$ solutions, we find the value of the 
central charge in the virasoro algebra: 
\beq\label{Rospella}
\frac{c}{12}= \frac{\phi_0 \beta A^{\frac{2(p-1)}{p+1}}p}{2^n}.
\feq
Our result for the central charge depends on the 
renormalization parameter $\beta$. The presence of this arbitrary
dimensionless constant 
is  a consequence of our renormalizations procedure.
From the point of view of   the 2D gravity theory, 
$\beta$ is  just a free parameter. However 
its  value can be 
constrained using arguments stemming from the AdS/CFT duality.
The central charge  in the Virasoro algebra  is a rational function of 
the conformal weights of 
 the boundary fields, so that we can expect  $\beta$ to be a rational 
number. Moreover, all the information about physical parameters of 
the $p$-brane is 
contained in the 2D parameters
 $A$ and $\phi_0$. The dimensionless parameter in Eq. (\ref{Rospella}) 
must  encode the information about the degrees of freedom 
of the $CFT_{p+1}$ living on the brane, leading again to a rational 
value for $\beta$.
We fix  
$\beta$, choosing  the value
\beq\label{Paola}
\beta=\frac{2^n}{p^2}.
\feq
The central charge takes the simple form,
\beq\label{RossellaAngius}
c=\frac{12}{p}\phi_0A^{\frac{2(p-1)}{p+1}}.
\feq
The entropy associated with the boundary CFT$_{1}$¥ characterized by  
 eigenvalue $l_0$ of the operator $L_0$ and central charge $c$, 
is given by the Cardy formula  
$S=2\pi \sqrt{c l_0/6}$, \cite{Cardy:ie}.
The  eigenvalue of $L_0$ is given in terms of 
the mass of the 2D black hole, $l_0=m_{bh}/b$, whereas 
$c$ can be read from  Eq. (\ref{RossellaAngius}).  We get for the 
entropy,
\beq\label{Angius}
S=2\pi\phi_0 A^{\frac{2p}{p+1}}.
\feq
Eq.(\ref{Angius})
holds  for all the non-dilatonic branes discussed in this paper.
Using Eqs. (\ref{Martedi4Maggio2004}, 
\ref{MTS}) to express $\phi_{0}$ and $A$ in terms 
of the brane temperature $T$ and  brane parameters $N,V,$
we reproduce exactly the thermodynamical entropy (\ref{entropy}),
$S_{p}¥=a_{p}V T^{p}$, with coefficients $a_{p}$ given by Eq. 
(\ref{a}).
By fixing appropriately the value of the renormalization parameter
$\beta$ our microscopical calculation of the brane entropy, which uses 
an effective AdS$_{2}$/CFT$_{1}$ duality, 
is in perfect agreement with the thermodynamical result. 

Notice that our general formula for the brane entropy (\ref{Angius}) 
holds also for $p=1$, although in this case no renormalization procedure, 
hence no fixing of the parameter $\beta$ is needed. 
In our 2D approach, the 1-brane (the BTZ black 
hole)  becomes after dimensional reduction   the AdS$_{2}$ black 
hole, whose microscopical entropy has been already calculated  in 
Ref. \cite{Cadoni:1998sg,Cadoni:1999ja,Caldarelli:2000xk,Cadoni:2000ah,
Cadoni:2000gm}. 

The weak point in our derivation is the fact that we do not have any 
compelling reason to fix $\beta$ as in Eq. (\ref{Paola}).
However, we can argue that Eq. (\ref{Paola}) may not be a simple 
coincidence. First, this value of $\beta$ 
seems to be rather special. With  this choice the central charge 
(\ref{RossellaAngius}) takes a simple form for all  
branes and the dimensionless factor in  the entropy  (\ref{Angius}) 
becomes  $p$-independent. Second, 
the factor $12/p$ appearing in the central charge  
(\ref{RossellaAngius}) seems related to the number of degrees of 
freedom of the $CFT_{p+1}$
living on the brane. The way how the information about CFT$_{p+1}$
degrees of freedom is encoded  in the  central charge of the CFT$_{1}$
may be  extremely non trivial.  However, our result seems to support 
recent attempts fo find  generalization of the Cardy formula for CFTs 
in $d>2$ \cite{Verlinde:2000wg}.

If we do not fix the renormalization parameter $\beta$, the entropy 
(\ref{Angius}) will depend on it. 
The dependence of the entropy from a dimensionless parameter can be 
also understood in terms of the classical scale symmetry of the 2D action
(\ref{Rossella}) mentioned in Sect. III. Rescaling the scalar field 
$\phi$, the  2D
action changes by an overall factor. 
This scale symmetry appears as a subgroup of the isometry group  of 
AdS$_{2}$. In fact, the metric (\ref{Pippo}) is invariant under the 
transformations
\beq\lb{st}
r\to\mu r,\quad t\to \mu^{-1} t.
\feq
This scale symmetry is broken by the scalar field, which encodes the 
information about the embedding of the brane in  the $D$-dimensional 
space-time. In fact  $\phi$ transforms as  $\phi\to \mu^{p}\phi$.
If we want to preserve the scale symmetry, the parameter $\phi_{0}$
must scale as $\phi_{0}¥\to \mu^{-p}\phi_{0}$.  Using this 
transformation law into Eq. (\ref{Angius}) we see that also the 
entropy scales in a similar way. This explains the dependence of the 
entropy  from a dimensionless parameter, which is undetermined, at 
least at the classical level.
Conversely, for the $AdS_{p+1}$ spacetime in Eq. (\ref{Nino}) the scale 
transformation (\ref{st}) can be promoted to an exact isometry 
transforming the brane coordinates $x^{i}\to \mu^{-1}x^{i}$.

\section{Conclusion}

In this paper  we have used a 2D approach  to study the microscopic
entropy of near-extremal non-dilatonic $p$-branes and, more in general,
to investigate the AdS/CFT correspondence  at finite temperature. 
Performing a dimensional reduction, we have found a 2D  gravity 
model that gives an effective description of the $p$-brane in the  
near-horizon, near-extremal regime. The AdS/CFT duality survives the 
dimensional reduction.  An AdS$_{2}$/CFT$_{1}$ duality gives an 
effective description of the AdS$_{p+2}$/CFT$_{p+1}$ correspondence at 
finite temperature. Finite temperature effects are taken into 
account in 
the 2D model as a breaking of the conformal symmetry, which generates 
a non-vanishing central charge in the Virasoro algebra. 
Using this procedure, we  have calculated the 
entropy of the boundary CFT$_{1}$. Fixing in a natural way a 
dimensionless free renormalization parameter, we have reproduced 
exactly the Bekenstein-Hawking entropy of all relevant non-dilatonic
$p$-branes in the near-extremal, near-horizon regime.

Our results represent an important  improvement, in particular for 
what concerns the 2- and 5-brane. For these branes methods based on 
the AdS$_{p+2}$/CFT$_{p+1}$ duality cannot explain the dependence of 
the entropy from the number of branes $N$. This is probably due to our 
lack of knowledge about M-theory and about the AdS$_{p+2}$/CFT$_{p+1}$
duality for $p=2,5$. Our 2D approach is more 
successful simply because it is almost completely based on 2D
gravitational physics, therefore largely independent from the details
of the fundamental theory in eleven dimensions.

On the other hand, the fact that a 2D model can be used as an 
unifying framework to describe all the relevant non-dilatonic  branes, 
indicates that the 2D description could be more general then it could 
seem at first sight. The reason behind this generality 
can be easily recognized. Similarly to what happens for black 
holes, also for black branes  the thermodynamical behavior is 
essentially
determined by the  2D $(r,t)$ sections of the spacetime and largely 
independent from the transverse dimensions.

The weakness of our 2D approach is that it is not  fully predictive. 
The microscopic entropy of the brane is determined up  to a 
dimensionless renormalization constant, which from the 2D point of 
view is a free parameter.  However, the values of this parameter 
that lead to agreement between statistical and thermodynamical 
entropy are natural from the point of view of the brane and seem to 
have an universal character. This may be the consequence  of the 
existence of  a general and deep
relationship between the central charge of the one-dimensional CFT
and the number of the degrees of freedom of the brane.



\begin{thebibliography}{99}

\bibitem{Maldacena:1997re}
J.~M.~Maldacena,
Adv.\ Theor.\ Math.\ Phys.\  {\bf 2} (1998) 231
[Int.\ J.\ Theor.\ Phys.\  {\bf 38} (1999) 1113]
[arXiv:hep-the/9711200].

\bibitem{Witten:1998qj}
E.~Witten,
Adv.\ Theor.\ Math.\ Phys.\  {\bf 2} (1998) 253
[arXiv:hep-the/9802150].


\bibitem{Gubser:1998bc}
S.~S.~Gubser, I.~R.~Klebanov and A.~M.~Polyakov,
Phys.\ Lett.\ B {\bf 428} (1998) 105
[arXiv:hep-th/9802109].



\bibitem{Gubser:1996de}
S.~S.~Gubser, I.~R.~Klebanov and A.~W.~Peet,
Phys.\ Rev.\ D {\bf 54} (1996) 3915
[arXiv:hep-th/9602135].

\bibitem{Klebanov:1996un}
I.~R.~Klebanov and A.~A.~Tseytlin,
Nucl.\ Phys.\ B {\bf 475} (1996) 164
[arXiv:hep-th/9604089].

\bibitem{Strominger:1997eq}
A.~Strominger,
JHEP {\bf 9802} (1998) 009
[arXiv:hep-th/9712251].

\bibitem{Strominger:1996sh}
A.~Strominger and C.~Vafa,
Phys.\ Lett.\ B {\bf 379} (1996) 99
[arXiv:hep-th/9601029].

\bibitem{Callan:1996dv}
C.~G.~.~Callan and J.~M.~Maldacena,
Nucl.\ Phys.\ B {\bf 472} (1996) 591
[arXiv:hep-th/9602043].


\bibitem{Horowitz:1996fn}
G.~T.~Horowitz and A.~Strominger,
Phys.\ Rev.\ Lett.\  {\bf 77} (1996) 2368
[arXiv:hep-th/9602051].

\bibitem{Breckenridge:1996sn}
J.~C.~Breckenridge, D.~A.~Lowe, R.~C.~Myers, A.~W.~Peet, A.~Strominger and C.~Vafa,
Phys.\ Lett.\ B {\bf 381} (1996) 423
[arXiv:hep-th/9603078].

\bibitem{Horowitz:1996ac}
G.~T.~Horowitz, D.~A.~Lowe and J.~M.~Maldacena,
Phys.\ Rev.\ Lett.\  {\bf 77} (1996) 430
[arXiv:hep-th/9603195].

\bibitem{Maldacena:1996ya}
J.~M.~Maldacena,
Nucl.\ Phys.\ B {\bf 477} (1996) 168
[arXiv:hep-th/9605016].

\bibitem{Cadoni:2003vi}
M.~Cadoni,
Class.\ Quant.\ Grav.\  {\bf 21} (2004) 251
[arXiv:hep-th/0306069].


\bibitem{Horowitz:cd}
G.~T.~Horowitz and A.~Strominger,
Nucl.\ Phys.\ B {\bf 360} (1991) 197.


\bibitem{Duff:1991pe}
M.~J.~Duff and J.~X.~Lu,
Phys.\ Lett.\ B {\bf 273} (1991) 409.

\bibitem{Duff:1994an}
M.~J.~Duff, R.~R.~Khuri and J.~X.~Lu,
Phys.\ Rept.\  {\bf 259} (1995) 213
[arXiv:hep-th/9412184].


\bibitem{Gregory:1995qh}
R.~Gregory,
Nucl.\ Phys.\ B {\bf 467} (1996) 159
[arXiv:hep-th/9510202].

\bibitem{Stelle:1996tz}
K.~S.~Stelle,
arXiv:hep-th/9701088.

\bibitem{Peet:1997es}
A.~W.~Peet,
Class.\ Quant.\ Grav.\  {\bf 15} (1998) 3291
[arXiv:hep-th/9712253].


\bibitem{Aharony:1999ti}
O.~Aharony, S.~S.~Gubser, J.~M.~Maldacena, H.~Ooguri and Y.~Oz,
Phys.\ Rept.\  {\bf 323} (2000) 183
[arXiv:hep-th/9905111].
%
\bibitem{Petersen:1999zh}
J.~L.~Petersen,
Int.\ J.\ Mod.\ Phys.\ A {\bf 14} (1999) 3597
[arXiv:hep-th/9902131].

\bibitem{Gibbons:1994vm}
G.~W.~Gibbons, G.~T.~Horowitz and P.~K.~Townsend,
Class.\ Quant.\ Grav.\  {\bf 12} (1995) 297
[arXiv:hep-th/9410073].


\bibitem{Gubser:1998nz}
S.~S.~Gubser, I.~R.~Klebanov and A.~A.~Tseytlin,
Nucl.\ Phys.\ B {\bf 534} (1998) 202
[arXiv:hep-th/9805156].


\bibitem{Klebanov:2000me}
I.~R.~Klebanov,
[arXiv:hep-th/0009139].

\bibitem{Danielsson:2001xe}
U.~H.~Danielsson, A.~Guijosa and M.~Kruczenski,
JHEP {\bf 0109} (2001) 011
[arXiv:hep-th/0106201].

\bibitem{Danielsson:2002qg}
U.~H.~Danielsson, A.~Guijosa and M.~Kruczenski,
Rev.\ Mex.\ Fis.\  {\bf 49S2} (2003) 61
[arXiv:gr-qc/0204010].

\bibitem{KalyanaRama:2004fk}
S.~Kalyana Rama,
[arXiv:hep-th/0404026].

\bibitem{Saremi:2004pi}
O.~Saremi and A.~W.~Peet,
[arXiv:hep-th/0403170].

\bibitem{Halyo:2004bx}
E.~Halyo,
[arXiv:hep-th/0406082].


\bibitem{Cadoni:2001ew}
M.~Cadoni, P.~Carta, M.~Cavaglia and S.~Mignemi,
Phys.\ Rev.\ D {\bf 65} (2002) 024002
[arXiv:hep-th/0105113].

\bibitem{Youm:1999xn}
D.~Youm,
Phys.\ Rev.\ D {\bf 61} (2000) 044013
[arXiv:hep-th/9910244].


\bibitem{Medved:2003ar}
A.~J.~M.~Medved,
Phys.\ Rev.\ D {\bf 69} (2004) 064023
[arXiv:gr-qc/0311071].


\bibitem{Cadoni:1998sg}
M.~Cadoni and S.~Mignemi,
Phys.\ Rev.\ D {\bf 59} (1999) 081501
[arXiv:hep-th/9810251].
%
\bibitem{Cadoni:1999ja}
M.~Cadoni and S.~Mignemi,
Nucl.\ Phys.\ B {\bf 557} (1999) 165
[arXiv:hep-th/9902040].
%
\bibitem{Cadoni:2000ah}
M.~Cadoni and S.~Mignemi,
Phys.\ Lett.\ B {\bf 490} (2000) 131
[arXiv:hep-th/0002256].
%
\bibitem{Cadoni:2000gm}
M.~Cadoni, P.~Carta, D.~Klemm and S.~Mignemi,
Phys.\ Rev.\ D {\bf 63} (2001) 125021
[arXiv:hep-th/0009185].
%
\bibitem{Caldarelli:2000xk}
M.~Caldarelli, G.~Catelani and L.~Vanzo,
JHEP {\bf 0010} (2000) 005
[arXiv:hep-th/0008058].
%
\bibitem{Astorino:2002bj}
M.~Astorino, S.~Cacciatori, D.~Klemm and D.~Zanon,
Annals Phys.\  {\bf 304} (2003) 128
[arXiv:hep-th/0212096].
%

\bibitem{Teitelboim:1972vw}
C.~Teitelboim,
Annals Phys.\  {\bf 79} (1973) 542.



\bibitem{Regge:1974zd}
T.~Regge and C.~Teitelboim,
Annals Phys.\  {\bf 88} (1974) 286.

\bibitem{Benguria:1976in}
R.~Benguria, P.~Cordero and C.~Teitelboim,
Nucl.\ Phys.\ B {\bf 122} (1977) 61.

\bibitem{Carlip:1999cy}
S.~Carlip,
Class.\ Quant.\ Grav.\  {\bf 16} (1999) 3327
[arXiv:gr-qc/9906126].


\bibitem{Carlip:2002be}
S.~Carlip,
Phys.\ Rev.\ Lett.\  {\bf 88} (2002) 241301
[arXiv:gr-qc/0203001].

\bibitem{Mann:1992yv}
R.~B.~Mann,
Phys.\ Rev.\ D {\bf 47} (1993) 4438
[arXiv:hep-th/9206044].

\bibitem{Cardy:ie}
J.~L.~Cardy,
Nucl.\ Phys.\ B {\bf 270} (1986) 186.


\bibitem{Verlinde:2000wg}
E.~Verlinde,
arXiv:hep-th/0008140.



\end{thebibliography}
\end{document}